\documentclass[aps,a4paper,twocolumn]{revtex4}
\usepackage{epsfig}
\usepackage{dcolumn}
\usepackage{times}
\usepackage{mathptm}

\begin{document}

% common abbreviations
\newcommand{\beq}[0]{\begin{equation}}
\newcommand{\eeq}[0]{\end{equation}}
\newcommand{\di}[0]{d}
\newcommand{\mulame}[0]{\mu_L}
\newcommand{\lamlame}[0]{\lambda_L}
\newcommand{\ups}[0]{\upsilon}
\newcommand{\half}[0]{{\textstyle \frac{1}{2}}}
\newcommand{\inv}[1]{{\textstyle \frac{1}{#1}}}
\newcommand{\breuk}[2]{{\textstyle \frac{#1}{#2}}}
\newcommand{\deriv}[2]{\frac{\partial #1}{\partial #2}}
\newcommand{\derivtwo}[2]{\frac{\partial^2 #1}{\partial {#2}^2}}

\title{Melting of Polydisperse Hard Disks}
\author{Sander Pronk} 
\author{Daan Frenkel}
\affiliation{
    FOM Institute for Atomic and Molecular Physics\\
    Kruislaan 407\\
    1098 SJ Amsterdam\\
    the Netherlands
}

\begin{abstract}
  The melting of a polydisperse hard disk system is investigated by
  Monte Carlo simulations in the semigrand canonical ensemble. This is
  done in the context of possible continuous melting by a dislocation
  unbinding mechanism, as an extension of the 2D hard disk melting
  problem. We find that while there is pronounced fractionation in
  polydispersity, the apparent density-polydispersity gap does not
  increase in width, contrary to 3D polydisperse hard spheres. The
  point where the Young's modulus is low enough for the dislocation
  unbinding to occur moves with the apparent melting point, but stays
  within the density gap, just like for the monodisperse hard disk
  system.  Additionally, we find that throughout the accessible
  polydispersity range, the bound dislocation-pair concentration is
  high enough to affect the dislocation unbinding melting as predicted
  by Kosterlitz, Thouless, Halperin, Nelson and Young.
\end{abstract}

\maketitle   

\section{Introduction}

In the 1930's, Landau and Peierls showed that two dimensional solids
are qualitatively different from their 3D counterpart, as they lack
long-ranged positional order (see e.g. Ref.~\cite{PeierlsSurprises}).
However, 2D crystals do have long-ranged bond-orientational order and,
in this respect, they differ form the isotropic liquid phase where
both translational and bond-orientational order are short ranged.

In the 1970's, Kosterlitz and Thouless suggested that melting of
two-dimensional crystals may be quite different form 3D melting.  In
particular, they proposed that melting in two dimensions may proceed
via a continuous dislocation-unbinding transition.

Kosterlitz and Thouless\cite{Nelson,ChaikinLubensky} showed that the
free energy associated with a single dislocation becomes negative when
\begin{equation}
  K < 16 \pi k_B T
  \label{p2d:ktpoint}
\end{equation}
where $K$ is the Young's modulus of the crystal.  At the point where
$K=16\pi k_BT$, dislocation pairs can unbind and, as solid with free
dislocations `flow' under shear, Kosterlitz and Thouless interpreted
this temperature as the melting point. In a more detailed analysis,
Halperin and Nelson\cite{Nelson} and Young\cite{YoungMelting2D},
showed that dislocation unbinding is not enough to complete the
melting process.  At the point where the condition of
Eq.~\ref{p2d:ktpoint} is first satisfied, the system undergoes a
(continuous) transition from a 2D crystal to a hexatic phase. The
hexatic phase is characterized by short-ranged (exponentially
decaying) positional order, but quasi-long-ranged (algebraically
decaying) orientational order: the positional order is destroyed by
the presence of the unbound dislocations. A second (continuous) phase
transition is required to transform the hexatic phase into an
isotropic liquid with short-ranged bond-orientational order.

The Kosterlitz-Thouless-Halperin-Nelson-Young (KTHNY) theory makes
precise predictions about the behavior of the correlation functions of
both the translational and orientational order parameters. It should
be stressed, however, that the KTHNY theory describes only a possible
scenario: it also possible that one or both of the continuous
transitions are first order, and even that there is a direct
first-order transition from the crystal to the isotropic fluid.

The KTHNY predictions sparked off an intensive search for real or
model systems that would exhibit this two-stage melting process (for
an early review see: ref.~\cite{Strandburg2DMelting}. More recent
examples can be found in
refs.~\cite{BagchiAndersenSwope96,ChenKaplanMostoller95,BladonFrenkel95}
and \cite{ZahnLenkeMaret99}).

Surprisingly, however, there is still no satisfactory answer to the
question whether the KTHNY scenario applies even to the simplest of
all two-dimensional model systems, namely hard, elastic disks. In
fact, this system was the very first to be studied in any computer
simulation~\cite{Metropolis1953}.

The reason why it is difficult to determine the nature of the
melting transition, is that finite size effects tend to obscure
the distinction between first order and continuous melting in 2D
systems (see ref.~\cite{Strandburg2DMelting}).

In the case of hard disks, the early work by Alder and Wainwright
suggested that the hard-disk melting transition was first order
\cite{AlderWainwrightDisks} (in the very early work of Metropolis et
al.~\cite{Metropolis1953}, the computing power was insufficient to draw
meaningful conclusions about the nature of the melting transition).
The hard-disk melting problem was revisited many times after the
suggestion of the KTHNY scenario, but the evidence is still ambiguous.
Evidence for continuous melting was reported in
ref.~\cite{FernandezContinuousMelting2D}, while evidence for a
first-order phase transition was presented in
refs.~\cite{LeeStrandburg2DHD, WojciechowskiHDElastic,
  WeberMarx,WeberMarxBinder}.
% H. Weber, D. Marx, K. Binder, Phys. Rev. B 51 (1995) 14636
% H.Weber, D. Marx, Europhys. Lett. 27 (1994) 593.
In addition, several publications could not distinguish between the
two scenarios~\cite{ZollwegChesterMelting2D,Jaster2DHDMelting,
  BatesFrenkelElasticDisks}. More recently, there has been some
evidence for the KTHNY
scenario\cite{SenguptaMelting2DHD,JasterShortTime2DHD,JasterHexatic}
%,JasterPreprint2003},
% A. Jaster cond-mat/0305239
but, the matter still seems far from settled.

One possible route to tackle this problem would be to consider
hard disks as a special case of a more general class of systems,
and study possible trends in the phase behavior of this
generalized model. In the present case, we consider the 2D
hard-disk system as a special case of polydisperse disks. In 3D
hard spheres, the melting transition is of first order. As the
polydispersity is increased the difference in volume fraction of
the coexisting solid and liquid phases widens with increasing
polydispersity\cite{BolhuisPdisp1,BolhuisPdisp2}. One might expect
similar behavior in two dimensions if the solid-liquid transition
would be of first order.

While polydispersity in two dimensional systems has been studied
before, it was in the context of melting by increasing size dispersity
for Lennard Jones systems \cite{SadrDispersityMelting} or in the
context of a possible glass
transition\cite{SantenAbsence,Santen2DPoly}. In the present paper, we
examine the phase behavior of polydisperse hard disk systems.

%\begin{figure}
  %\center
  %\includegraphics[width=4.0cm]{poly2D/dislocation}
  %\includegraphics[width=4.0cm]{poly2D/dislocationpair}
  %\caption{Left: a dislocation as a pair of disclinations
    %(particles with 5 and 7 neighbours in this case) on a triangular
    %lattice. Right: a tightly bound disclocation pair.  Both pictures
    %are from actual simulations (with greatly reduced particle
    %diameter). The triagulation is a Delaunay triangulation.}
  %\label{p2d:dislocation}
%\end{figure}

\section{The System}

\subsection{The Semigrand Canonical Ensemble}

The model for polydispersity is based on the semigrand canonical
ensemble\cite{KofkeSemigrand}; this ensemble has previously been used
to study the phase diagram of polydisperse 3D hard
spheres\cite{BolhuisPdisp1,BolhuisPdisp2}. The semigrand canonical
ensemble can be seen as a hybrid version of the canonical ensemble and
the grand canonical ensemble. It is characterized by a thermodynamic
potential $X$ that satisfies the following fundamental thermodynamic
relation\cite{Callen}
\begin{equation}
  \di X = - S \di T - P \di V - \int
  N(\sigma) \delta \mu(\sigma) \di \sigma
\end{equation}
Here $S$ is the entropy of the system, $T$ the temperature, $P$
the pressure, $V$ the volume. $N(\sigma)d\sigma$ denotes the
number of particles with diameter between $\sigma$ and
$\sigma+d\sigma $ and $\mu(\sigma)$ is the chemical of particles
with diameter $\sigma$. We now add and subtract a term containing
the chemical potential of a reference species $\sigma_0$
%\begin{equation}
%  \int N(\sigma) \delta \mu(\sigma_0)  \di \sigma
%    =
%     \di \mu(\sigma_0) \int N(\sigma)\di \sigma
%    = N \di \mu(\sigma_0)
%\end{equation}
from the complete differential:
\begin{equation}
  \di X = -S \di T - P \di V - N \di\mu(\sigma_0)
  - \int
  N(\sigma) \delta \Delta \mu(\sigma) \di \sigma
\end{equation}
where we replace $\mu(\sigma)-\mu(\sigma_0)$ with $\Delta
\mu(\sigma)$. We now perform a Legendre transformation to a new
ensemble that has $N$ as a thermodynamic control parameter instead
of $\mu(\sigma_0)$ (and $P$ instead of $V$).
\begin{equation}
  \di Y = -S \di T + V \di P + \mu(\sigma_0)\di N
  - \int
  N(\sigma) \delta \Delta \mu(\sigma) \di \sigma
\end{equation}
which, in explicit form becomes (with the Euler equation)
\begin{eqnarray}
  Y(N,\Delta \mu(\sigma),P,T) & = &
  U - TS + PV + N \mu(\sigma_0)
  \nonumber \\
  & & - \int N(\sigma) \Delta \mu(\sigma) \di \sigma
  \nonumber \\
  & = & N \mu(\sigma_0)
\end{eqnarray}

The partition sum for this ensemble is
\begin{eqnarray}
  \Upsilon(N,\Delta \mu(\sigma),P,T) =
  \int \di \sigma^N \int \di V \int \di \mathrm{s}^N
  \nonumber \\
  \exp\left( -\beta \left\{PV + U(V,\mathrm{s}^N) -
    \Delta \mu(\sigma) N(\sigma) ] \right\} \right)
\label{p2d:partsum}
\end{eqnarray}
with $Y = -k_BT \ln \Upsilon$. For simulation purposes, the
semigrand canonical ensemble can be interpreted as one where there
is constant number of particles that can change identity. This
identity switching can be an extra move in a Monte Carlo
simulation; in a polydisperse mixture, this would amount to a
particle size change with an acceptance criterion based on the
functional form for the  chemical potential $\Delta \mu(\sigma)$.

As in Ref. \cite{BolhuisPdisp1}, we use the following functional
form for the chemical potential
\begin{equation}
  \Delta \mu(\sigma) = - k_BT \frac{(\sigma - \sigma_0)^2}{2\nu^2}
  \label{p2d:mufunctional}
\end{equation}
which, at zero density, will give a Gaussian particle size
distribution according to the partition sum of Eq.~\ref{p2d:partsum}.
In practice, the size distribution is Gaussian-like at the densities
of the crystalline phase.

\subsection{Order Parameters}

For the study of the properties of the phases and the phase
transitions, some order parameters have been used that are standard in
the studying of 2D melting\cite{Strandburg2DMelting} and for which the
KTHNY melting scenario makes explicit predictions\cite{Nelson}. We
define the $n$-fold bond-orientational order at $\mathbf{x}_i$, the
position of particle $i$, as
\begin{equation}
  \psi_n(\mathbf{x}^i) =  \frac{1}{N_i} \sum_{j=1}^{N_i}
  e^{in\theta_{j}(\mathbf{x}^i)}
\end{equation}
where $N_i$ is the number of neighbors and $\theta_{j}(\mathbf{x}^i)$
is the angle between an arbitrary (fixed) axis and the line connecting
particle $i$ with its $j$-th neighbor; two particles are neighbors if
they share a Voronoi cell edge. For systems that tend to crystallize
into triangular lattices, the leading bond-order parameter is the one
for which $n=6$.  The global value of the order parameter is simply
the mean of the local values.

The positional order is measured using the static structure factor
$S(\mathbf{q})$ at one specific scattering vector $\mathbf{q}$
equal to a reciprocal lattice vector of a perfect crystal with
orientation and lattice spacing taken from the system. To check
for hexagonal crystalline positional order, the lattice vector
$a_0$ is set to its ideal value for a given packing fraction:
\begin{equation}
  a_0 = \left( \frac{\pi/\sqrt{12}}{\eta} \right)^{1/2}
\end{equation}
The crystal orientation is taken from the mean angle obtained from the
global hexagonal bond-orientational order parameter:
\begin{equation}
  \Psi_6= \frac{1}{N} \sum_{i=1}^N \psi_6(\mathbf{x}_i)
  \label{p2d:psi6global}
\end{equation}
which specifies the orientation of one of the six equivalent
crystal axes within an angular range $0 \le \alpha < \pi/3$. Once
the average orientation of the nearest-neighbor `bonds' ${\bf
a_0}$ has been specified, it is straightforward to deduce the
orientation of the corresponding reciprocal lattice vector
$\mathbf{G}$ through $\mathbf{G} \cdot {\bf a_0}=2 \pi$. The
positional order parameter $\zeta$ of the crystal is then given by
\begin{equation}
  \zeta(\mathbf{x}_i) = e^{i \mathbf{G} \cdot \mathbf{x}_i}
\end{equation}
Radial correlation functions of the order parameters are defined
as
\begin{eqnarray}
  g_6(r)& = & %\langle \exp[i\sum \theta_{ij} - \sum \theta_{kl}] \rangle 
   \langle \psi_6^*(\mathbf{0}) \psi_6(r) \rangle / g(r)
  \\
  \zeta(r)& =&\langle \zeta^*(\mathbf{0}) \zeta(r) \rangle / g(r)
\end{eqnarray}
In two-dimensional systems, $\zeta$ is expected to decay to zero,
either exponentially (short ranged order) or algebraically
(`quasi-long ranged'). The KTHNY melting scenario makes predictions
for the decay of the orientational order in the hexatic and phase:
$g_6(r) \sim r^{-\eta_6}$ with $\eta_6 \to \frac{1}{4}$ at the melting
of the hexatic phase into the liquid
phase\cite{Nelson,Strandburg2DMelting}.

\subsection{Elasticity}
\label{p2d:elasticity}

Eq.~\ref{p2d:ktpoint} provides a very useful test to decide whether or
not a 2D melting can be of the KTHNY type. If we find that Young's
modulus drops below the `magical' value of $16\pi k_BT$ in an
otherwise stable solid, then it is very likely that this solid melts
by dislocation unbinding. Conversely, if we find that this magical
value is only crossed at densities where we know that the isotropic
liquid phase is thermodynamically stable, then it is reasonable to
assume that melting is a first-order transition.  Often, however, the
simulations do not provide a clear answer, as the point where $K=16\pi
k_BT$ is located in the intermediate density regime that may either be
a two-phase region separating two stable phase, or the domain of the
elusive hexatic phase.

The Young's modulus is defined through the shear ($\lamlame$) and bulk
($\mulame$) Lam\'e elastic constants in 2D,
\begin{equation}
  \label{p2d:young}
  K=\frac{4a_0^2\mulame (\mulame + \lamlame)}{2 \mulame + \lamlame} \le 16 \pi
\end{equation}
where $a_0$ is the equilibrium lattice spacing. The Lam\'e elastic
constants are related to the second-order elastic constants (i.e. the
elastic constants defined by the second derivative of the free energy
to the Lagrangian strain), through
\begin{eqnarray}
  C_{11} & = & \lamlame + 2 \mulame    \nonumber \\
  C_{12} & = & \lamlame                \nonumber \\
  C_{44} & = & \mulame - P
\end{eqnarray}

The theory of the KTHNY scenario is based on a group renormalization
of both the Young's modulus and the dislocation fugacity
\begin{equation} 
y=exp^{-\beta E_c} 
\end{equation} 
where $E_c$ is the core energy of a dislocation. As the length scale
of the renomalization is increased towards infinity, the dislocation
fugacity and the elastic constants that constitute $K$ tend to their
`renormalized' values, where the (long-range) effect of (presumably
low-concentration) dislocations is properly taken into account.

The elastic constants measured in this work, although calculated at
finite sizes, should be close to their renormalized values because of
the high concentration of dislocations in the hard disk system, as
will be shown in Section\ref{p2d:simsec}.

When specifying elastic constants of a polydisperse system, we should
distinguish between `quenched' and `annealed' elastic constants.
Quenched elastic constants measure the second strain derivative of the
free energy of a polydisperse crystal with a `frozen' size
distribution: i.e. the particle-size distribution is assumed not to
respond to the deformation. In contrast, the `annealed' elastic
constants measure the second strain derivative of the semi-grand
potential.  In this case, the particle size distribution is assumed to
respond to the deformation. The quenched constants are the ones that
are presumably measured in mechanical experiments that probe the
elastic deformation of a polydisperse solid. But the annealed
constants describe the equilibrium state of a deformed polydisperse
solid. In order to determine the critical value for Young's modulus in
a polydisperse 2D solid, we computed the annealed elastic constants,
as these determine the equilibrium behavior of the system.

To calculate the elastic constants at different polydispersities, a
hybrid Monte Carlo -- Molecular Dynamics method was used, where Monte
Carlo runs sampling particle diameters and positions in the strained
system were followed by Molecular Dynamics runs where the stress
tensor was measured using the instantaneous realization of the
particle size distribution. The algorithm thus effectively calculates
the elastic constants of many realizations of a polydisperse
hard-sphere crystal.

The quenched elastic constants %described in section \ref{hsel:polydisp} 
will be larger than the annealed quantities, because of the concavity
of the free energy. Hence the corresponding `quenched' Young's modulus
will only reach the instability limit $K=16\pi$ at lower densities.

The strains used in the elastic constant determination were
\begin{eqnarray}
  \alpha^1_{ij} & = & \left( \begin{array}{cc} 1+a & 0 \\ 0 & 1+a
    \end{array} \right)
    \nonumber \\
  \alpha^2_{ij} & = & \left( \begin{array}{cc} 1 & b \\ 0 & 1
    \end{array}  \right)
\end{eqnarray}
with $a$ as the (small) strain parameter. The resulting stress
derivatives %,  according to Eq.~\ref{intro:stressstrain}, 
give us\cite{WallaceElastic}
\begin{eqnarray}
\frac{d T_{11}}{d a} & = & \frac{\partial T_{11}}{\partial
  \alpha^1_{11}} + \frac{\partial T_{11}}{\partial
  \alpha^1_{22}} = C_{11} + C_{12} = 2\lamlame + 2\mulame
\nonumber \\
\frac{d T_{12}}{d b} & = & \frac{\partial T_{12}}{\partial
  \alpha^1_{12}} = C_{44} + P = \mulame
\end{eqnarray}
allowing us to calculate the Young's modulus of Eq.~\ref{p2d:young}.

\section{Simulations}
\label{p2d:simsec}

\begin{figure}
  \center
  \includegraphics[width=0.95\linewidth]{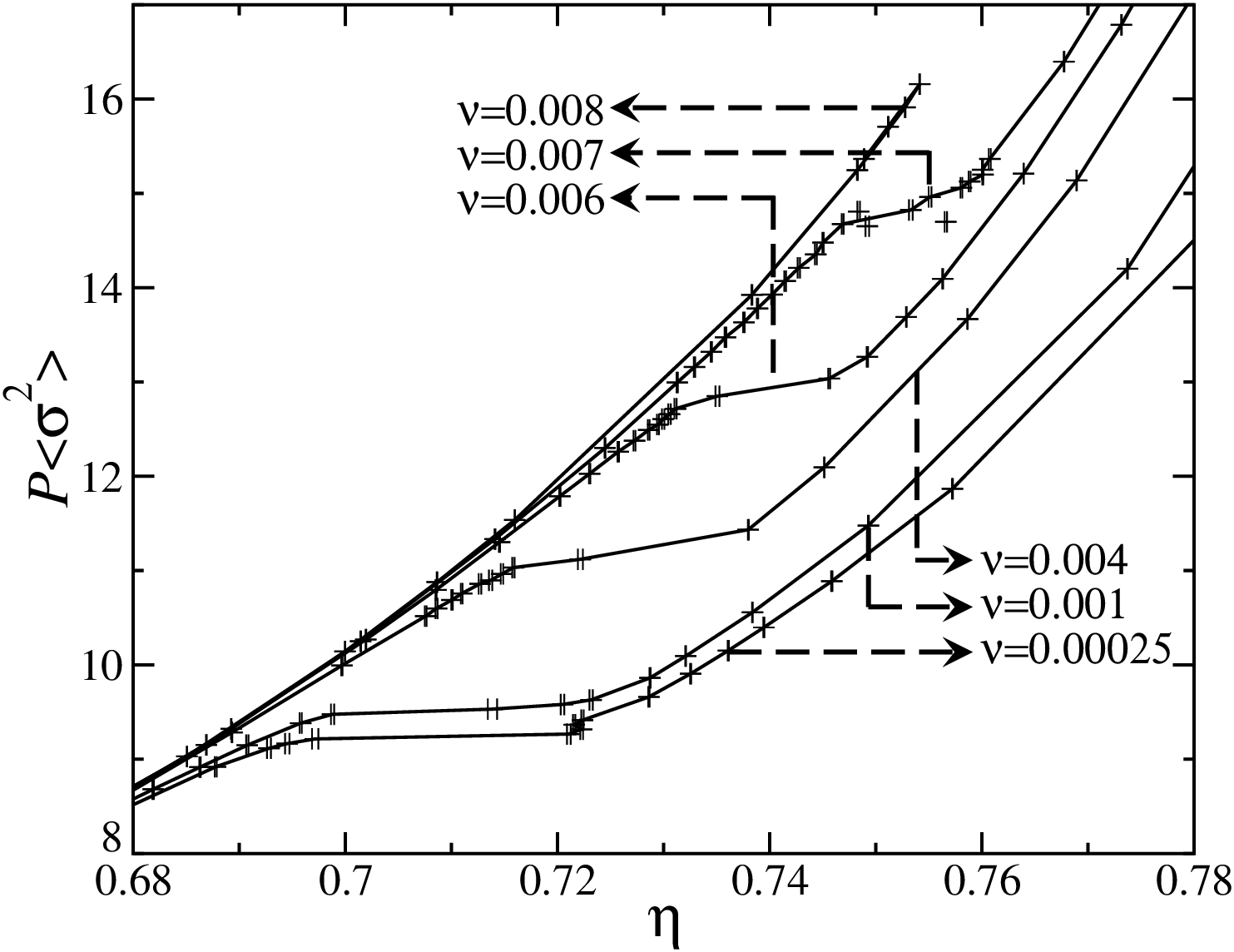}
  \caption{Calculated equations of state (with normalized pressure) for
    varying polydispersity control parameter $\nu$. System size is
    $4012$ particles.}
  \label{p2d:eqst}
\end{figure}

The simulations were performed on systems containing $59\times68=4012$
particles, with a chemical potential disitribution width
(`polydispersity') parameter $\nu$ (see Eq.~\ref{p2d:mufunctional})
varying between $\nu=0.00025$ and $\nu=0.008$. For higher values of
$\nu$ (higher polydispersities), the equilibration was exceedingly
slow, even with the combined volume--particle radius sampling:
equilibration took $1.3 \times 10^{6}$ to $2.5 \times 10^6$ MC steps
per particle and data were sampled during $1\times10^6$ steps per
particle; around $20\%$ of steps were devoted to particle radius
steps. For $N=4012$, the simulation time was 6--8 hours on an Athlon
1600+ CPU.

Monte Carlo simulations of the semigrand canonical ensemble of
Eq.~\ref{p2d:partsum} involve displacement moves, volume moves and identity
change moves, which in the current context mean particle radius moves. 
To speed up the simulation, the coupling between mean particle radius
and system size is exploited to remove the global volume moves and
simultaneous system+particle size scaling was introduced. This
improves the statistics of sampling considerably\cite{BolhuisPdisp1}.
It turns out that, in practice, a real simultaneous system+particle
size scaling is at least as fast as the analytical integration of the
scaling part of the partition sum also described in
Ref.~\cite{BolhuisPdisp1}.

The resulting equations of state are shown in Fig.~\ref{p2d:eqst}.
From this figure, it is clear that the phase transition, which is
around packing fraction $\eta \approx 0.70$ for the monodisperse
case\cite{Jaster2DHDMelting,BatesFrenkelElasticDisks,ZollwegChesterMelting2D},
shifts to higher packing fractions and higher pressures with
increasing polydispersity. At $\nu=0.008$, the system cannot be made
to freeze at all. A similar phenomenon was observed in
ref.~\cite{BolhuisPdisp1} --- it is due to the choice of the
functional form for the chemical potential
(Eqn.~\ref{p2d:mufunctional}).  In addition, for $\nu=0.007$, it is
extremely difficult to equilibrate the system properly in the vicinity
of the phase transition.

\begin{figure}
  \center
  \includegraphics[width=0.95\linewidth]{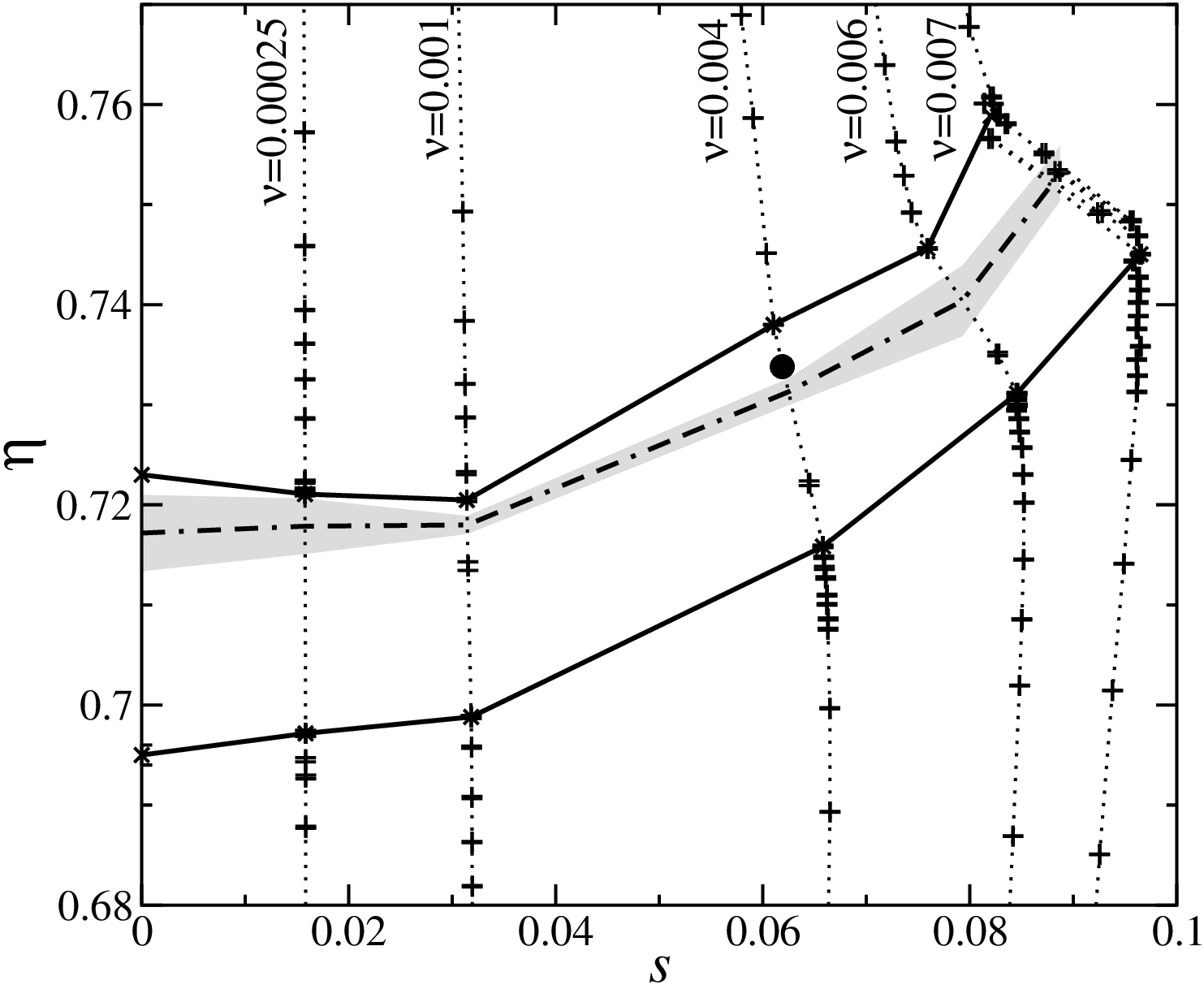}
  \caption{Packing fraction $\eta$ as a function of polydispersity $s$ 
    for varying $\nu$, with $N=4012$. The boundaries of the apparent
    density gap are shown as the solid lines; the density gap of the
    monodisperse ($s=0$) case is taken from data by
    Jaster\cite{Jaster2DHDMelting}, with $N=4096$. The dashed line
    shows the location of packing fractions (with matching
    polydispersities) extrapolated to be $K=16\pi$; the gray area
    denotes an estimate of the size of the error. The black circle
    shows the location of the $K=16\pi$ point obtained using a larger
    system size ($N=4012$) for the elasticity simulations.}
  \label{p2d:poly}
\end{figure}

\begin{figure}
  \center
  \includegraphics[width=0.95\linewidth]{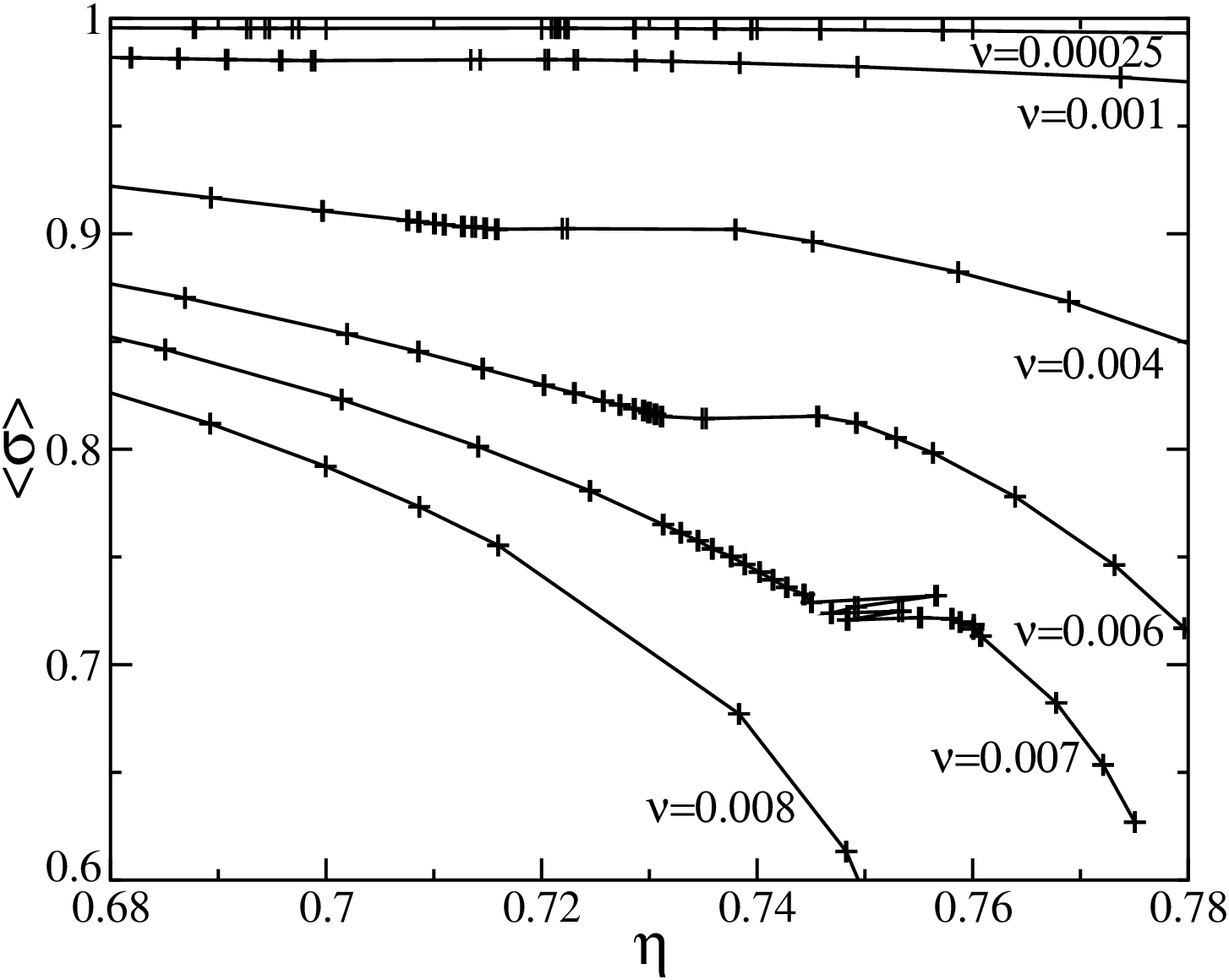}
  \caption{Mean particle diameter $\langle\sigma\rangle$ as a function
    of packing fraction $\eta$ for the different values of the polydispersity
    control parameter $\nu$.}
  \label{p2d:etasigma}
\end{figure}

As is the case for 3D spheres, the liquid branches of the
equations of state very nearly superimpose~\cite{BolhuisPdisp1}.
However, near the phase transition, the pressure appears to
increase slightly with polydispersity. Upon further compression,
the system undergoes a phase transition with an apparent density
gap. However, because of the relatively small system size, the
presence of such an apparent density gap is also compatible with
continuous melting. We find that the density gap decreases with
increasing polydispersity.

In Fig.~\ref{p2d:poly} we plot the variation in polydispersity upon
freezing. As a measure for the polydispersity, we use $s \equiv
\langle \sigma^2 \rangle/\langle \sigma \rangle^2 -1$.  In the same
figure, the apparent density gap is also shown (the borders of the
density gap shown here are simply the solid with the lowest packing
fraction and the liquid with the highest packing fraction found in the
simulations). As in the 3D case~\cite{BolhuisPdisp1,BolhuisPdisp2},
the freezing point moves to higher volume fractions as the
polydispersity is increased and size fractionation is also increased.
But, whereas the density gap widens upon freezing in 3D, it appears,
if anything, to decrease in 2D (except for the possibly seemingly
poorly equilibrated case of $\nu=0.007$). Additionally, the maximum
polydispersity at which the solid seems to remain stable (around 8\%)
is considerably higher than in 3D ($5.7 \%$).

In the semigrand-canonical ensemble, the system equilibrates to a mean
particle size, dependent on pressure and $\nu$.
Fig.~\ref{p2d:etasigma} shows this mean particle size as a function of
packing fraction: the phase transitions show up as jumps in the
packing fraction, but not in the mean particle size. This means that
there is no particle size fractionation (the mean of the particle
diameter distribution does not change) while there is, as can be seen
in Fig~\ref{p2d:poly}, polydispersity-fractionation (the width of the
particle diameter distribution changes).

\begin{figure}
  \center
  \includegraphics[width=0.47\linewidth]{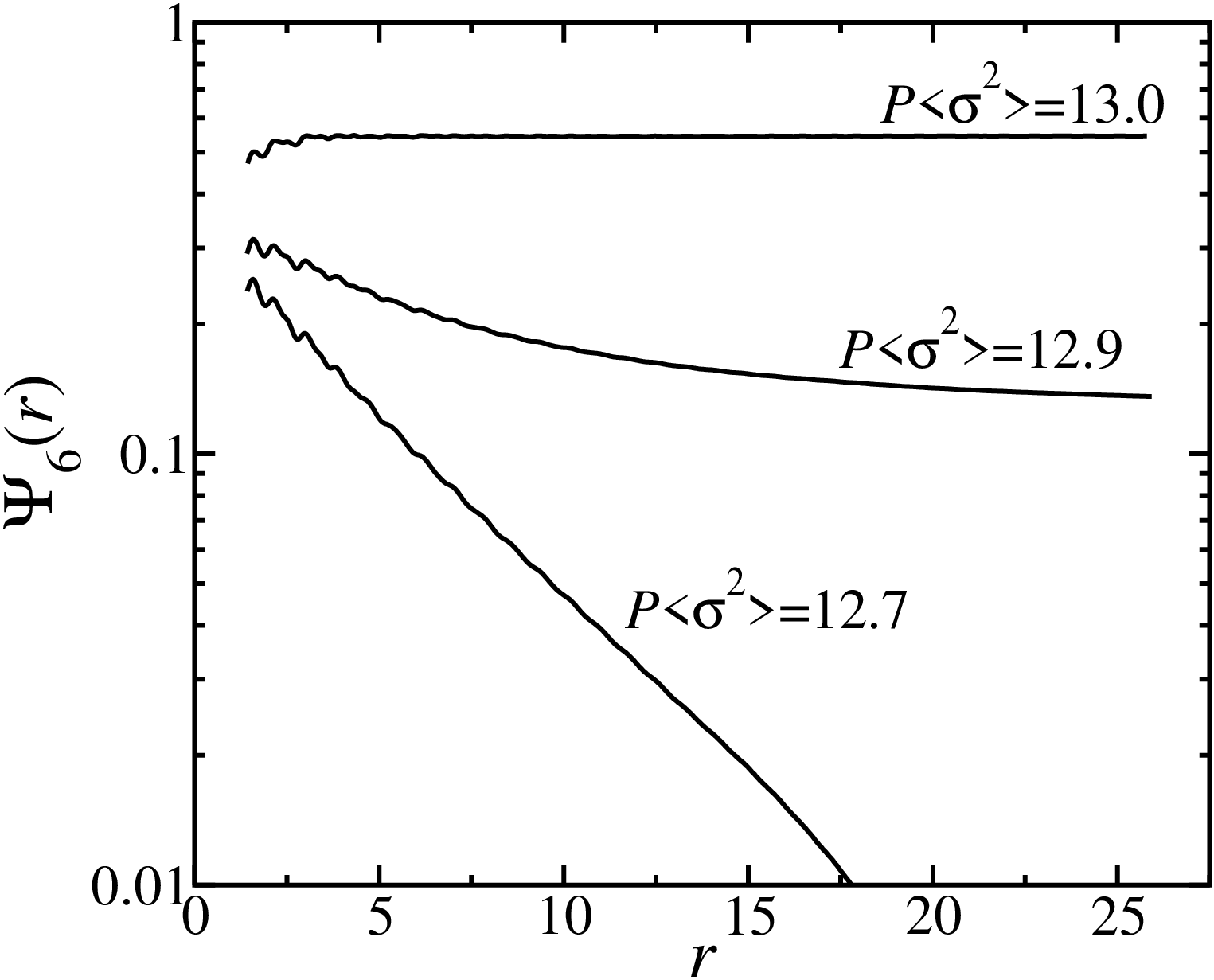}
  \hspace{0.02\linewidth}
  \includegraphics[width=0.47\linewidth]{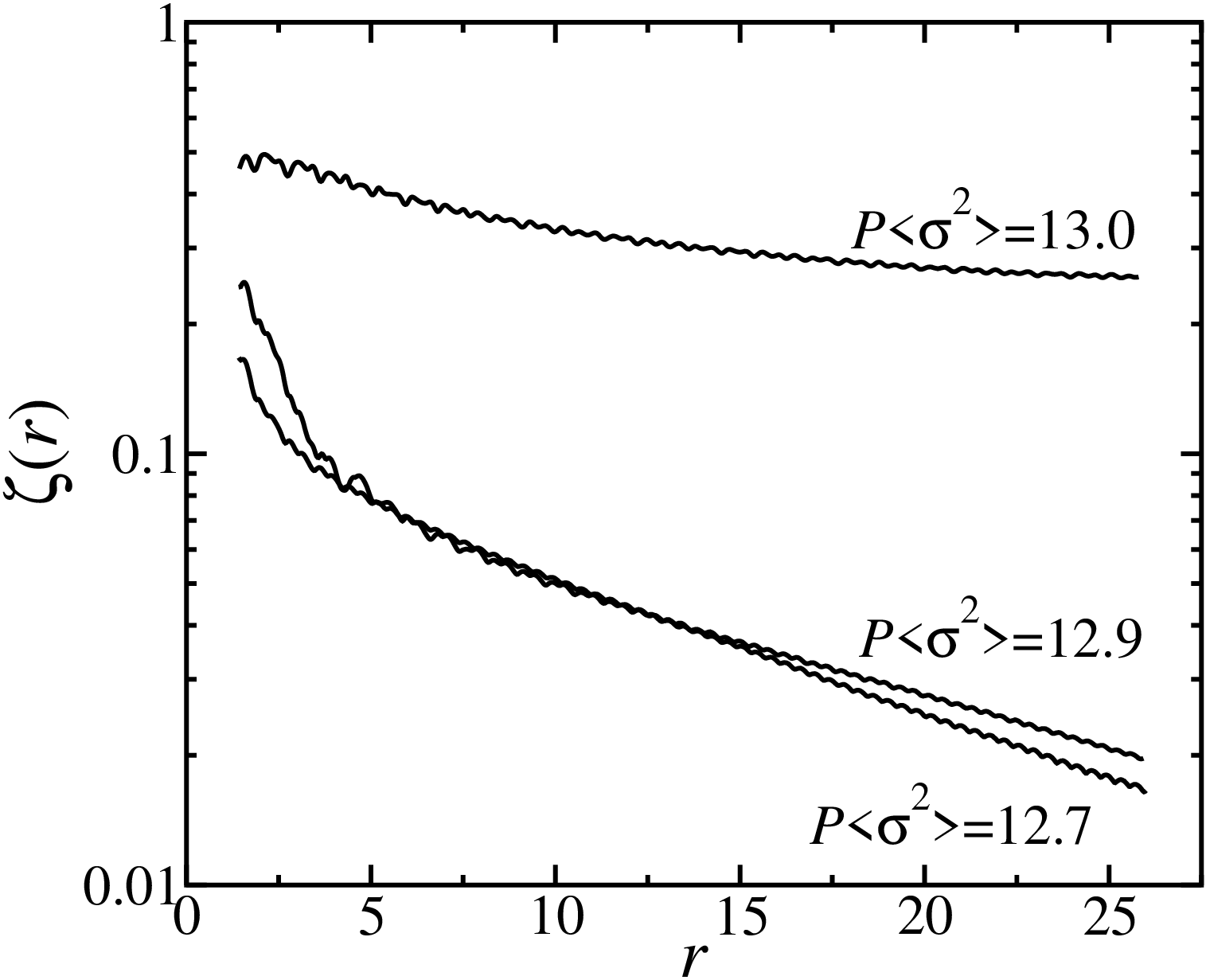}
  \caption{Example of the decay of the orientational (left) and positional
    (right) order parameters over the phase transition, for
    $\nu=0.006$. The decay of the positional order parameter,
    $g_6(r)$ goes as $g_6(r)\propto r^{-\frac{1}{4}}$ at the
    intermediate pressure $P\langle\sigma^2\rangle=12.7$. The
    positional order parameters decay exponentially for both lower
    pressures.}
  \label{p2d:order}
\end{figure}

An example of the behavior of the order parameters near the phase
transition is shown in Fig.~\ref{p2d:order}. The positional and
orientational order both increase sharply (but not quite
simultaneously) in the region of the phase transition, similar to what
one finds for monodisperse hard disks. Although the decay of the
orientational correlation function $g_6(r)$ goes with
$r^{-\frac{1}{4}}$ as predicted by the KTHNY scenario for the hexatic
at the hexatic-crystal transition, this seems to be a coincidence,
because a further simulation of the same configuration yields a
different (but still algebraic) decay rate.  This indicates that the
typical decay `time' of fluctuations in the system exceeds the typical
length of the (rather long) simulations.

\begin{figure}
  \center
  \includegraphics[width=0.95\linewidth]{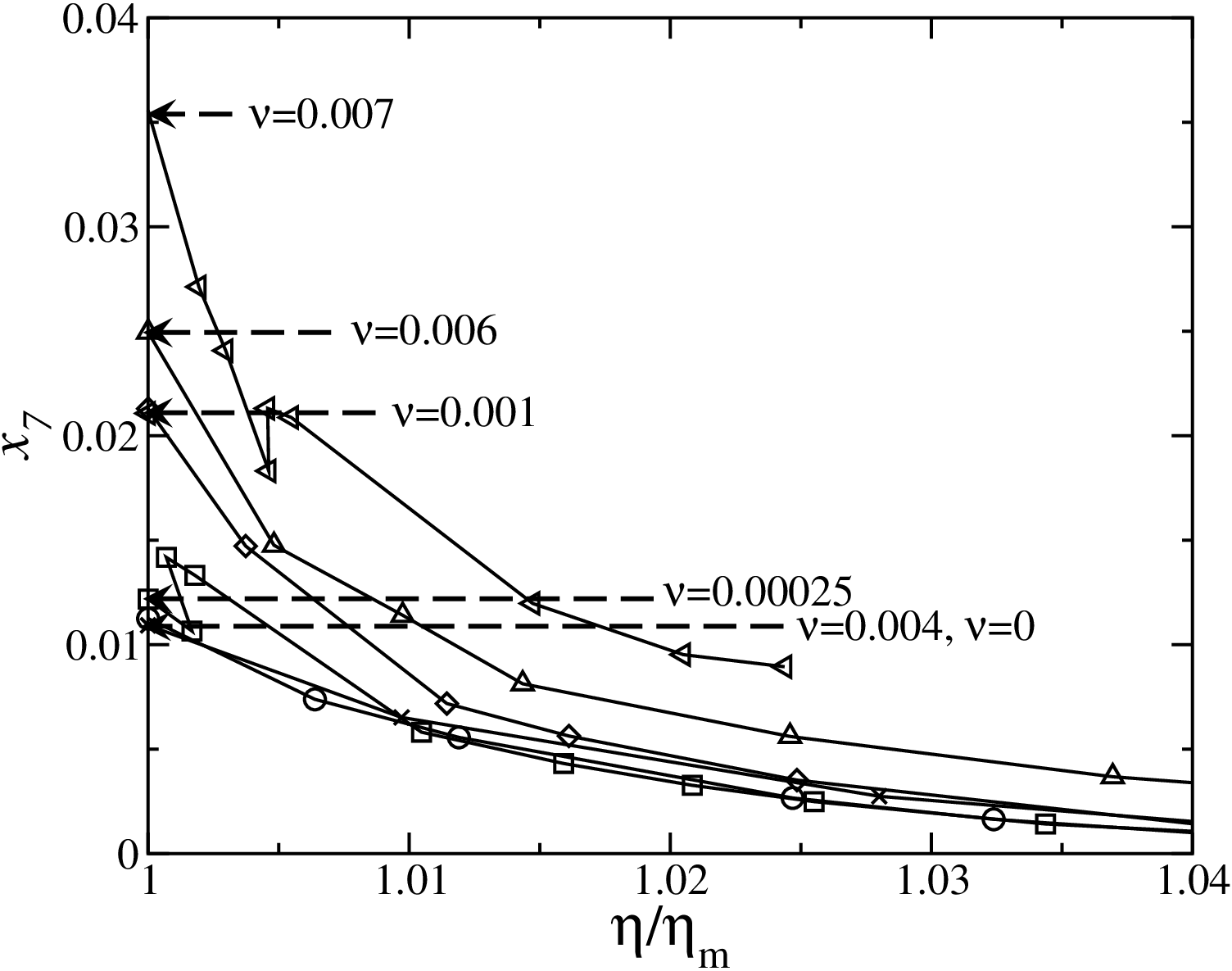}
  \caption{Concentration of seven-coordinated particles $x_7$ (a
    measure for the number of dislocations) as a function of packing
    fraction relative to the melting packing fraction (determined by
    the boundaries of the apparent density gap).}
  \label{p2d:sevennbr}
\end{figure}

We stress that all observations reported thus far are consistent with
either a KTHNY scenario or a weak first-order transition.  The KTHNY
scenario however, is based on the assumption that the concentration of
bound dislocations in the crystal phase is low; dislocation
interaction is not taken into account and the Kosterlitz-Thouless
normalization is based on an expansion in the dislocation unbinding
length which may be unrealistically long for high dislocation
concentrations. We measured the concentration of bound dislocation
pairs (see Fig.~\ref{p2d:sevennbr}). In the figure, we show the
concentration of seven-coordinated particles. Because in the crystal
the number of eight or more coordinated particles is negligible, and
as the number of point defects turns out to be an order of magnitude
lower than the number of bound
dislocations\cite{BatesFrenkelElasticDisks}, the number of
seven-coordinated particles is a good measure of the dislocation
count.  At the melting point (the boundary of the apparent density
gap) the dislocation concentration varies from 1\% to more than 3\%.
As the concentration of dislocation pairs depends sensitively on the
dislocation core energy, this suggests that the core energy is rather
low. Note that a core energy less than $2$ to $4 k_B T$ is not
compatible with KTHNY
melting\cite{SaitoDislocationVector,ChuiGrainBoundaryMelting}.

Neglecting the dislocation-dislocation interaction (a very crude
approxximation), but including the internal elastic energy and entropy
of dislocation pairs\cite{Fisher79a,SenguptaMelting2DHD}, this defect
density would be consistent with a core energy of approximately
$5k_BT$ over the full polydispersity range.  This, however, is a crude
approximation and the real core energy may be several $k_BT$ off.  The
value is compatible with both the value of $6 k_BT$ given in
Ref~\cite{SenguptaMelting2DHD} and the value of $11 k_BT$ at a much
higher packing fraction of $0.82$ calculated in Ref.~\cite{Bladon04a}.

\section{Elastic Constants}

The elastic constants were measured using the method described in
section \ref{p2d:elasticity}, using simulations of $412$ particles,
equilibrating for $4\times10^6$ MC steps and $4\times10^4$ MD
collisions per particle and measuring up $5\times10^6$ MC steps and
$1.5\times10^6$ MD collisions per particle.  Earlier
work~\cite{BatesFrenkelElasticDisks} had shown that the elastic
constants are not significantly affected by the presence of point
defects such as vacancies.  Away from the KTHNY transition, the
elastic constants are not very sensitive to finite-size
effects~\cite{WojciechowskiHDElastic}. Of course, this is not true
close to a KTHNY transition, but this will turn out to be less
relevant for the present system because we always observe melting
before we get into the `danger zone'. 

\begin{figure}
  \center
  \includegraphics[width=0.95\linewidth]{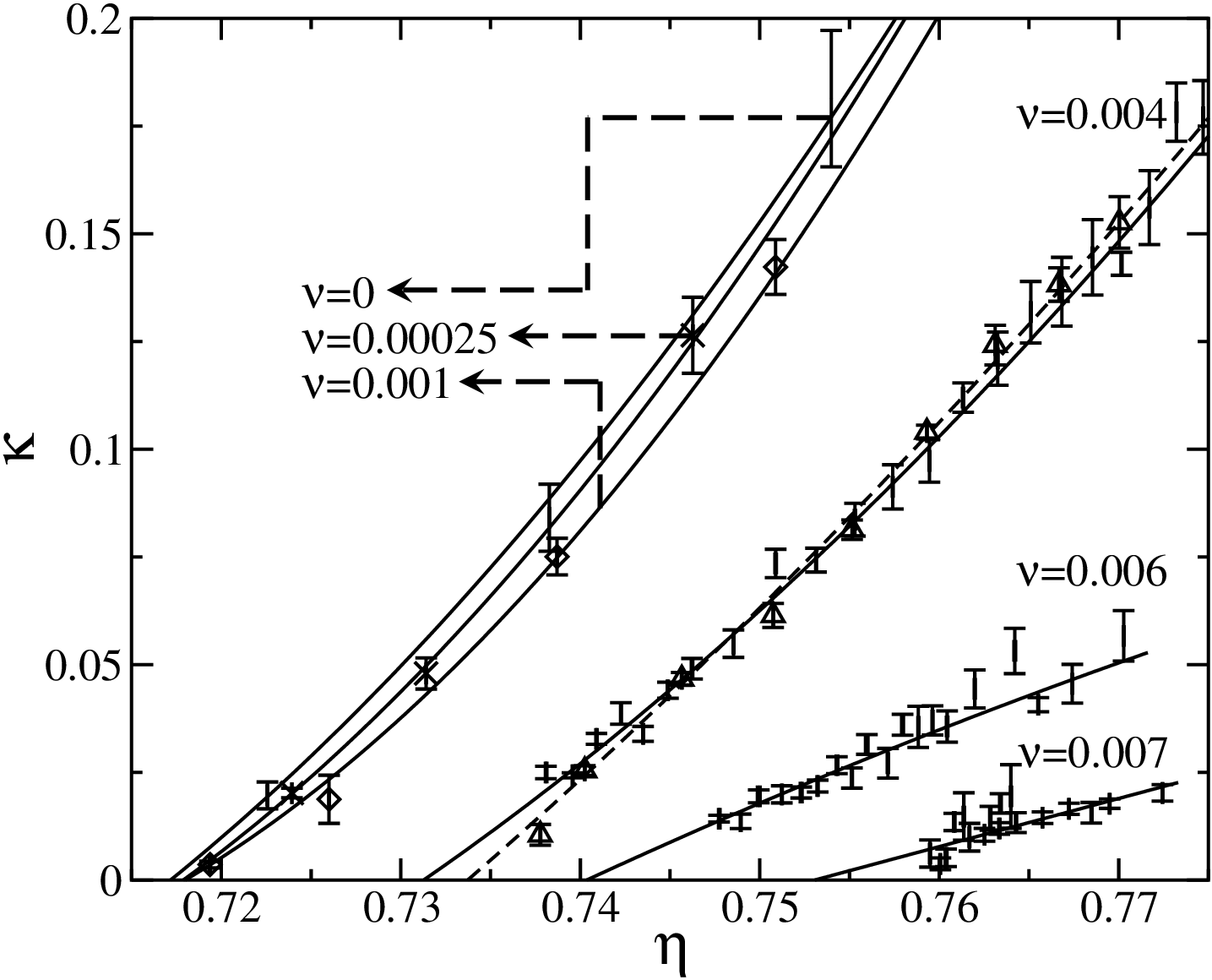}
  \caption{Fractional exponent of the Young's modulus
    $\kappa=(1-16\pi/K)^{\nu_{\mathrm{KT}}}$ with
    $\nu_{\mathrm{KT}}=0.3696$ () for the different polydispersities,
    as a function of packing fraction. The lines are the fits to
    determine the locations of $K=16\pi$. The points at $\nu=0.004$
    with the triangles together with the fit shown with the dashed
    line are results of calculations done with 4012 particles.  
    }
  \label{p2d:phtr-K}
\end{figure}

The results of the simulations are shown in Fig.~\ref{p2d:phtr-K};
here an exponential of the Young's modulus,
$\kappa=(1-16\pi/K)^{\nu_{\mathrm{KT}}}$, where
$\nu_{\mathrm{KT}}$ is the Kosterlitz-Thouless renormalization
exponent, which has a value of $\nu_{\mathrm{KT}} \approx
0.36963$\cite{Nelson}. This form is chosen because close to the KTHNY
transition, the Young's modulus --- expressed as function of the
difference between the packing fraction $\eta$ and the packing
fraction at the KTHNY, $\eta_{\mathrm{KT}}$ --- behaves
as\cite{Strandburg2DMelting}
\begin{equation}
  \frac{K}{16\pi} = 
  \frac{1}{1-c(\eta-\eta_{\mathrm{KT}})^{\nu_{\mathrm{KT}}}}
\end{equation}
The measured $\kappa$ values are then fitted to a second-order
polynomial. The results for the monodisperse case ($\nu=0$) are
similar to those by Wojciechowski and Bra\'{n}ka
\cite{WojciechowskiHDElastic,WojciechowskiHDClosePackedElastic} and by
Bates and Frenkel\cite{BatesFrenkelElasticDisks}. 

To check for finite-size effects in the elastic constants
calculations, the calculations for $\nu=0.004$ were also done with
4012 particles. The results are shown in Fig.~\ref{p2d:phtr-K} and in
Fig.~\ref{p2d:poly}; the $K=16\pi$ extrapolation is very similar to
the smaller 412 particle system and agrees within the error margin.

From the locations of the $K=16\pi$ line in Fig.~\ref{p2d:poly} it is
clear that the situation with respect to the type of phase transition
is, even for higher polydispersities, similar to that of the
monodisperse system; the points seem to follow --- within the
statistical uncertainties --- not only the loci of the phase
transitions, but also the position within the density gap.

\section{Conclusion}

In this article, we explored the effect of polydispersity on the
nature of the 2D melting transition is a system of 2D hard disks.

We find that the solid-liquid phase transition shifts to higher
packing fractions as the polydispersity increases, and that
polydispersity fractionation takes place in the region of the phase
transition.  The maximum polydispersity at which the solid can be
stable is larger than in 3D hard spheres.

The density-polydispersity gap, be it real or apparent, does not seem
to increase in size with increasing polydispersity. The fact that the
points for which $K=16\pi$ appear to be located in the two-phase
region, supports the assumption that the melting transition is first
order. Even if this should not be the case, the high dislocation
concentration will presumably have an effect on the KTHNY predictions.

The work of the FOM institute is part of the research program of the
Foundation for Fundamental Research on Matter (FOM) and was made
possible through financial support by the Dutch Foundation for
Scientific Research (NWO).

\end{document}